\documentstyle[epsf]{mn}

\begin{document}

\title[Be/X-ray binaries in the Magellanic Clouds]
    {The optical counterparts to Be/X-ray binaries in the Magellanic Clouds}

\author[J.B.Stevens, M.J.Coe and D.A.H.Buckley]
{J.B.Stevens$^{1}$, M.J.Coe$^{1}$, D.A.H.Buckley$^{2}$\\
$^{1}$Dept. of Physics \& Astronomy, University of Southampton,
Southampton, SO17 1BJ, UK \\
$^{2}$South African Astronomical Observatory, P.O. Box 9, Observatory,
7935, South Africa
}

\date{Accepted \\
Received : Version \today\\
In original form ..}

\maketitle

\begin{abstract} 

The fields of 8 X-ray sources in the Magellanic Clouds believed to be
Be/X-ray binaries have been searched for possible Be star
counterparts. $BVR_{c}$ and H$\alpha$ CCD imaging was employed to
identify early type emission stars through colour indices and
H$\alpha$ fluxes.  Spectroscopy of 5 sources confirms the presence of
H$\alpha$ emission in each case. Based on the positional coincidence
of emission line objects with the X-ray sources, we identify Be star
counterparts to the {\it ROSAT} sources RX J0032.9-7348, RX
J0049.1-7250, RX J0054.9-7226 and RX J0101.0-7206, and to the recently
discovered {\it ASCA} source AX J0051-722. We confirm the Be star
nature of the counterpart to the {\it HEAO1} source H0544-66. In the
field of the ROSAT source RX J0051.8-7231 we find that there are three
possible counterparts, each showing evidence for H$\alpha$
emission. We find a close double in the error circle of the {\it
EXOSAT} source EXO 0531.1-6609, each component of which could be a Be
star associated with the X-ray source.

\end{abstract}


\section{Introduction}

The Magellanic Clouds (MC's) present a unique opportunity to study
stellar populations in galaxies other than our own. Their structure
and chemical composition differs from that of the Galaxy, yet they are
close enough to allow study with modest sized ground based
telescopes. The study of any stellar population in an external galaxy
is of interest as any differences with the same population in our own
Galaxy will have implications on the evolutionary differences of the
stars within the galaxies. A High Mass X-ray Binary (HMXB) consists of
a compact object (neutron star or black hole) in orbit around a
non-degenerate massive (OB type) star. X-ray emission is the result of
accretion of material onto the compact object from the massive
companion. The HMXB's can be divided into those with supergiant
companions, and those with Be star companions. A Be star is defined as
an early type luminosity class III--V star which has at some time
shown emission in the Balmer lines. This Balmer emission, along with a
significant infrared excess is believed to originate in the
circumstellar material which forms a disc around the star. The X-ray
emission in these systems is transient in nature, as the orbit is wide
and eccentric, and the neutron star passes through the densest regions
of the circumstellar disc at periastron only. In the supergiant
systems, the companion fills or is close to filling its Roche lobe,
and mass transfer occurs through Roche lobe overflow or via a strong
stellar wind removing $\sim$10$^{-8}$M$_{\odot}$y$^{-1}$. These
systems tend to be more persistent than the Be/X-ray binaries,
sometimes showing flaring events on short timescales (For a
comprehensive review of X-ray binaries, see Lewin, van Paradijs \& van
den Heuvel 1995.)\par Observations of the HMXB's in the Magellanic
Clouds appear to show marked differences in the populations. The X-ray
luminosity distribution of the MC sources appears to be shifted to
higher luminosities relative to the Galactic population. There also
seems to be a higher incidence of sources suspected to contain black
holes (see Clark et al. 1978; Pakull 1989; Schmidtke et
al. 1994). Clark et al. (1978) attribute the higher luminosities to
the lower metal abundance of the MC's, whilst Pakull (1989) refers to
evolutionary scenarios of van den Heuvel \& Habets (1984) and de Kool
et al (1987) which appear to favour black hole formation in low metal
abundance environments.\par

In order to study the differences between the HMXB populations of the
Magellanic Clouds and the Galaxy, it is desirable to determine the
physical parameters of as many systems as possible. We can then
investigate whether the distributions of mass, orbital period, or
spectral type are significantly different. Because of the small sample
size of known Be/X-ray binaries in the Magellanic Clouds we have
searched the fields of a number of unidentified X-ray sources
suspected to be HMXB's in an attempt to identify more Be/X-ray
binaries.

Table 1 lists the X-ray sources in the Magellanic Clouds observed
during this study. The sample was chosen to include unidentified X-ray
sources from which either pulsations have been detected, or other
characteristics that strongly suggests a HMXB
status.

\begin{table}
\centering
\caption{Optically unidentified X-ray sources (suspected to be HMXB's)
observed from the SAAO. The quoted uncertainties are all 90\% confidence.}
\begin{tabular}{l c c c}
\hline
Source & RA  & Declination  & Uncertainty \\
&(J2000) & (J2000) & radius(arcsec)\\
\hline
RX J0032.9-7348 & 00 32 55.1 & -73 48 11 & 62 \\
RX J0049.1-7250 & 00 49 04.6 & -72 50 53 & 22 \\
AX J0051-722    & 00 50 55.8 & -72 13 38 & 10 \\ 
RX J0051.8-7231 & 00 51 53.0 & -72 31 45 & 11 \\
1WGA J0054.9-7226&00 54 56.1 & -72 26 45 & 11 \\
RX J0101.0-7206 & 01 01 03.2 & -72 06 57 & 10  \\
EXO0531.1-6609  & 05 31 12.0 & -66 07 08 & 9\\
H0544-665       & 05 44 15.5 & -66 33 50 & 30\\
\hline
\end{tabular}
\end{table}

In Table 1, column 4 gives the uncertainty in X-ray position in
arcseconds. Many of the sources are {\it ROSAT} detections with
uncertainties of the order of a few arcseconds. In galactic fields,
sub-ten arcsecond resolution would normally be adequate for an
unambiguous optical identification, but owing to the crowded nature of
Magellanic Cloud fields, even some of the {\it ROSAT} sources have
X-ray positional uncertainties that allow several possible optical
ccandidates.

We obtained CCD images of the fields through BV(R)$_{C}$ and H$\alpha$
filters in order to identify early type stars within the fields, and
to search for H$\alpha$ emission from these stars. We have also
obtained medium- and low-resolution spectroscopy of most candidates in
order to confirm the presence of H$\alpha$ emission, and to measure
radial velocities to allow confirmation of SMC or LMC membership.

In the following sections we describe in more detail the observations
and subsequent analysis, and present resulting data for identified
candidates.

\section{Observations}

\begin{figure*}
%\begin{center}
%{\epsfxsize 0.81\hsize
%\leavevmode
%\file{finders.tif}
%$}\end{center}
%\vspace{11cm}
\caption{Finding charts (3$\times$3 arcmin): (a) RX J0032.9-7348; (b)
RX J0049.1-7250; (c) AX J0051-722; (d) RX J0051.8-7231; (e) 1WGA
J0054.9-7226; (f) RX J0101.0-7321; (g) EXO 0531.1-6609; (h) H
0544-665. In the case of charts (a) - (c) and (e) - (g) the error
circle shown comes from ROSAT observations. In the case of chart (d)
the larger circle is the ROSAT observations of Kahabka \& Pietsch
(1996), whereas the smaller circle is from the ROSAT observations of
Israel et al. (1997). The circle in chart (h) is from HEAO-1. Detailed
references to all the satellite observations may be found in the
text. All images are R band images from this work except charts (c),
(d) and (h) which have been produced from the Digitised Sky Survey
data. This figure is quite large and a compressed PS file (size 680K) may be
obtained from www.astro.soton.ac.uk/$\sim$mjc/images/fig1.ps.Z }
\label{finders}
\end{figure*}

\subsection{CCD Photometry}

CCD imaging was performed at the SAAO during 1996 October 1--7. All
observations were made using the 1.0-m telescope and Tek8 CCD, plus 3x
Shara focal reducer. The resulting pixel scale was 1.05'' per pixel,
with a total image size of 519x519''. All fields were observed through
R and H$\alpha$ filters. In addition, most fields were observed
through B and V filters. The H$\alpha$ filter used was an
interference filter centered on 6562\AA, with a width of 50\AA. A
complete log of observations is shown in Table~\ref{photlog} and
finding charts for all the targets are presented in Figure 1.

The use of the focal reducer, whilst necessary to provide a field of
view adequate to search larger X-ray error circles, introduced
significant vignetting which was not satisfactorilly removed by
flat-field corrections. Analysis showed that flat-field errors were
below the 1\% level within 4' of the image centre. In subsequent
analysis we therefore rejected any measurements of objects that lay
further than 4' from the image centre. 

Due to the crowded nature of the fields, profile fitting photometry
was necessary. PSF fitting photometry was carried out using
IRAF/DAOPHOT. In each field between 30 and 50 stars were used to model
the PSF. Instrumental magnitudes were transformed to the standard
system using observations of a set of E and F region standards. 

A H$\alpha$ magnitude scale was calibrated by defining the zero point
such that a R--H$\alpha$ index has a value of zero for main sequence,
non-emission line stars, and becomes positive for emission
line stars.

For each field observed, an {\it emission-colour} diagram was plotted,
with the R-H$\alpha$ emission index on the vertical axis, and the
B-V colour index on the horizontal axis. As demonstrated by Grebel
(1997), when such a diagram is plotted for Magellanic Cloud
fields, where all objects lie at the same distance and are affected by
the same reddening, Be stars can be clearly distinguished by their
blue colour, and high emission index. Comparing photometry of other
objects observed during this run with photometry obtained on other
occasions without use of the focal reducer indicated that errors of up
to 0.1 in magnitude could have resulted from fitting a poorly sampled
PSF. In addition, some observations were made in conditions which were
slightly below photometric quality, introducing systematic errors. As
the method of identification of Be star candidates is through their
relative positions in the $emission-colour$ diagram, systematic errors
do not undermine our results.

\begin{table}
\caption{Log of CCD imaging observations made with the 1.0m telescope
at the SAAO in 1996 October.}
\begin{tabular}{l c c c}
\hline
Source & Date & Filter band & Exposure (secs)\\
\hline
RX J0032.9-7348 & October 1 & B         &  200\\
                &           & V         &  200\\
                &           & R         &  200\\
                &           & H$\alpha$ & 1000\\
RX J0051.8-7231 & October 1 & R         &  150\\
                &           & H$\alpha$ & 1000\\
RX J0101.0-7206 & October 1 & R         &  150\\
                &           & H$\alpha$ & 1000\\
H0544-665       & October 1 & B         &  200\\
                &           & V         &  200\\
                &           & R         &  100\\
                &           & H$\alpha$ & 1000\\
RX J0049.1-7250 & October 2 & B         &  300\\
                &           & V         &  300\\
                &           & R         &  100\\
                &           & H$\alpha$ & 1000\\
RX J0054.9-7226 & October 5 & B         &  300\\
                &           & V         &  200\\
                &           & R         &  200\\
                &           & H$\alpha$ & 1000\\
\hline
\end{tabular}
\label{photlog}
\end{table}

\subsection{Optical spectroscopy}

\begin{table*}
\caption{Log of spectroscopic observations made with the 1.9m
telescope at the SAAO (all in 1998)}
\begin{tabular}{l c c c c c}
\hline
Object & Date & Exposure & Grating & Wavelength range & Dispersion\\
\hline
AX J0051-722    & Feb 3 & 900+900 & 5 & 6295 - 7042~\AA& 0.42~\AA/pixel\\
RX J0101.0-7201 & Feb 3 & 900+900 & 5 & 6295 - 7042~\AA& 0.42~\AA/pixel\\
EXO 0531.1-6609 & Feb 3 & 900+900 & 5 & 6295 - 7042~\AA& 0.42~\AA/pixel\\
		& Feb 5 & 600     & 7 & 3800 - 7779~\AA& 2.2~\AA/pixel\\
H0544-665       & Feb 3 & 900+900 & 5 & 6295 - 7042~\AA& 0.42~\AA/pixel\\
		& Feb 5 & 600     & 7 & 3800 - 7779~\AA& 2.2~\AA/pixel\\
RX J0032.9-7348 & Feb 4 & 900+1200& 5 & 6295 - 7042~\AA & 0.42~\AA/pixel\\
RX J0051.8-7231 & Feb 4 & 900+1200& 5 & 6295 - 7042~\AA & 0.42~\AA/pixel\\
RX J0054.9-7226 & Feb 4 & 900+900 & 5 & 6295 - 7042~\AA & 0.42~\AA/pixel\\
                & Feb 5 & 300+600 & 7 & 3800 - 7779~\AA & 2.2~\AA/pixel\\
\hline
\end{tabular}
\label{speclog}
\end{table*}

Optical spectra were obtained with the 1.9-m telescope at the SAAO and
the Cassegrain spectrograph with SITe 1 CCD. On 1998 February 3 and 4
spectra were obtained with a spectral range of 6295--7042~\AA~ and
dispersion of 0.43~\AA/pixel. On 1998 February 5, low resolution
spectra were obtained with a range of 3800--7780~\AA, and a dispersion
of 2.3~\AA/pixel. A log of spectroscopic observations is shown in
Table~\ref{speclog}. All spectra were reduced using tasks in IRAF's
KPNOSLIT package. Spectra were optimally extracted, and wavelength
scales were applied using arc-lamp spectra obtained before and after
each target observation.

On 1998 February 5, a flux standard was observed, as well as a smooth
spectrum standard for the removal of telluric features. All low
resolution spectra have subsequently been flux calibrated, and
telluric features to the red of H$\alpha$ have been removed. 

\section{Results and discussion of individual sources}

\subsection{RX J0032.9-7348} 

\begin{figure*}
\begin{center}
{\epsfxsize 0.9\hsize
\leavevmode
%\epsffile{r0032_bvrh.eps}
\epsffile{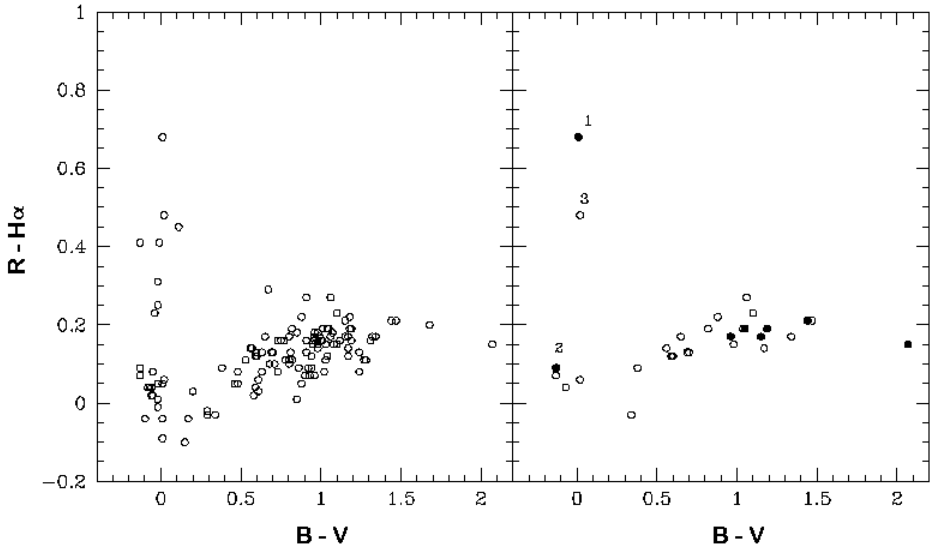}
}\end{center}
%\vspace{11cm}
\caption{{\bf Left:} H$\alpha$ emission index versus $B-V$ colour index for
objects in the field of the X-ray source RX
J0032.9-7348. {\bf Right:} The same diagram for objects within 124
arcseconds (twice the positional uncertainty) of the ROSAT X-ray
position. Objects lying within the error circle are plotted with
filled circles.}
\label{0032bvrh}
\end{figure*}

\begin{figure*}
%\begin{center}
%{\epsfxsize 0.81\hsize
%\leavevmode
%\epsffile{ha.eps}
%}\end{center}
%%\vspace{11cm}
\caption{H$\alpha$ spectra of the optical counterparts to the X-ray
sources.(a) RX J0032.9-7348; (b) AX J0051-722; (c) RX J0051.8-7231;
(d) 1WGA J0054.9-7226; (e) RX J0101.0-7321; (f) EXO0531.1-6609; (g) H0544-665.}
%\label{rxj0032_ha}
\end{figure*}

%\begin{table}
%\begin{centering}
%\caption{Photometry of the objects within the X-ray error
%circle of RX J0032.9-7348. Only those objects detected in all of $BVR$
%and H$\alpha$ frames are included.}
%\begin{tabular}{lccccc}
%\hline
%& V & B-V &b-v &R-H$\alpha$\\
%\hline
%1 & 13.62$\pm$0.04  & 0.59$\pm$0.06 & 0.39$\pm$0.05& 0.12$\pm$0.06\\
%2 & 15.11$\pm$0.04  & -0.13$\pm$0.06& -0.14$\pm$0.05&0.10$\pm$0.06\\
%3 & 16.26$\pm$0.05  & 0.01$\pm$0.07 & 0.04$\pm$0.06 &0.68$\pm$0.06\\
%4 & 16.69$\pm$0.05  & 1.05$\pm$0.07 & 0.70$\pm$0.06 &0.18$\pm$0.06\\
%5 & 16.13$\pm$0.05  & 2.07$\pm$0.10 & 1.38$\pm$0.08 &0.14$\pm$0.10\\
%6 & 15.14$\pm$0.04  & 1.10$\pm$0.06 & 0.84$\pm$0.05 &0.22$\pm$0.06\\
%7 & 14.36$\pm$0.04  & 0.60$\pm$0.06 & 0.39$\pm$0.05 &0.12$\pm$0.06\\
%8 & 17.16$\pm$0.06  & 0.96$\pm$0.08 & 0.57$\pm$0.07 &0.17$\pm$0.08\\
%9 & 15.79$\pm$0.05  & 1.15$\pm$0.08 & 0.70$\pm$0.06 &0.16$\pm$0.07\\
%\hline
%\end{tabular}
%\label{0032tab}
%\end{centering}
%\end{table}

This source was discovered by Kahabka \& Pietsch (1996, hereafter
KP1996) in {\it ROSAT} pointed observations made in 1992 December and
1993 April. The unabsorbed bolometric luminosity they derive from the
1993 observations is $2.5\times10^{36}$ ergs s$^{-1}$, whilst the 1992
December flux was a factor of $\sim$6 less. From the X-ray spectrum,
length of the X-ray high state (at least 5 days in 1993 April), and
long term variability, KP1996 propose a likely HMXB nature for the
source. The X-ray position was determined to an accuracy of
$\pm$~62~arcseconds(KP1996).

We obtained $BVR$ and H$\alpha$ images of the field on the night of
1996 October 1. Figure~\ref{0032bvrh} shows the $emission-colour$ diagram
for stars in the field, the left hand plot showing all stars in the
field, the right hand plot showing only those stars within a field
centered on the X-ray position, with a radius 124 arcseconds (twice
the positional uncertainty). On this plot, we further identify those
stars that lie within the X-ray error circle by plotting with filled
circles. The field population is predominantly evolved red stars, with
only six early type stars detected within the 124 arcsecond area. Of
these early type stars, two show clear excess H$\alpha$ flux. The
strongest H$\alpha$ excess is seen from a star which lies within the
X-ray error circle ($\sim$12 arcseconds from the X-ray position,
marked as Object~1 in Figures 1(a) and ~\ref{0032bvrh}). The other H$\alpha$
source (object~3 in Figure~\ref{0032bvrh}) lies $\sim$100 arcseconds
to the north of the X-ray position. Besides object~1, only one other
object within the error circle is identified as an early type star
from its $B-V$ colour (object~2, which we identify with
GSC~0914101338). 
%Table~\ref{0032tab} lists the photometry for the 9
%objects within the X-ray error circle detected in each of the $BVR$ and
%H$\alpha$ frames.

Our medium resolution spectrum of object~1 (Fig 3(a)) shows H$\alpha$ in emission
with an equivalent width of EW(H$\alpha$)~=~-35~\AA. No He I (6678\AA)
feature is seen above the level of the continuum noise. The H$\alpha$
line is single peaked, and centered on $6566.5\pm0.5$~\AA. Assuming that the
deviation from the rest wavelength of H$\alpha$ is purely due to the
radial velocity of the star, we derive a velocity of
$171\pm23$~km~s$^{-1}$, consistent with the systemic radial velocity
for the SMC of $166\pm3$~km~s$^{-1}$ found by Feast (1961). This
object is the most probable optical counterpart for RX J0032.9-7238
(but see the discussion on chance probability in Section 4).

\subsection{RX J0049.1-7250}

%\begin{figure*}
%\begin{center}
%{\epsfxsize 0.9\hsize
%\leavevmode
%\epsffile{r0049_bvrh.eps}
%}\end{center}
%\vspace{11cm}
%\caption{{\bf Left:} H$\alpha$ emission index versus $b-v$ colour index for
%objects detected in each of the $BVR$ and $H\alpha$ images of the
%field of the X-ray source RX J0049.1-7250. {\bf Right:} Same diagram
%only for objects within 44 arcseconds (twice the positional
%uncertainty) of the ROSAT X-ray position. Objects that lie within the
%X-ray error circle are plotted as filled circles.}
%\label{rxj0049bvrh}
%\end{figure*}

%\begin{table}
%\begin{centering}
%\caption{Photometry of the objects within the X-ray error
%circle of RX J0049.1-7250. Only those objects detected in all of $BVR$
%and H$\alpha$ frames are included.}
%\begin{tabular}{lccccc}
%\hline
%& V & B-V &b-v &R-H$\alpha$\\
%\hline
%1 &17.24$\pm$0.03&  0.05$\pm$0.04& -0.07$\pm$0.07& -0.02$\pm$0.08\\
%2 &17.40$\pm$0.04& -0.01$\pm$0.05& -0.03$\pm$0.07& -0.03$\pm$0.10\\
%3 &16.28$\pm$0.06&  0.08$\pm$0.08&  0.02$\pm$0.08&  0.32$\pm$0.07\\
%4 &17.12$\pm$0.05&  0.79$\pm$0.10&  0.46$\pm$0.06&  0.14$\pm$0.05\\
%\hline
%\end{tabular}
%\label{0049tab}
%\end{centering}
%\end{table}

This source was discovered with {\it ROSAT} (KP1996) in pointed
observations. It appears highly absorbed, and is variable by a factor
of more than 10. KP1996 concluded that the source probably lay behind
the SMC, with a maximum luminosity of $\sim10^{38}$~ergs, but they
could not rule out a time variable background AGN nature for the source.

A new X-ray pulsar was discovered by the {\it RXTE} satellite during
observations centered on the position of SMC X-3. Pulsations were
detected with a period of 74.8$\pm$0.4~seconds (Corbet et
al. 1998b). Follow up observations with the {\it ASCA} satellite made on
1997 November 13 detected pulsations with a period of
74.675$\pm$0.006~seconds (Yokogawa \& Koyama 1998). The 2
arcminute X-ray error circle from the ASCA observations contained the
{\it ROSAT} error circle for RX J0049.1-7250. Kahabka \& Pietsch (1998)
report that the source showed a high degree of variablility, having
gone undetected in a 35~ksec {\it ROSAT} HRI observation on 1997 May 9-25,
and concluded that a Be/X-ray binary nature was probable for the
source.

A finder chart is shown in Figure 1(b) with the ROSAT error circle
marked. An {\it emission-colour} diagram similar to Figure 2 was
produced for this field. Only one Be star is found within the X-ray
error circle, lying only 3~arcseconds from the X-ray position (marked
as object~1 in Fig 1(b). A second Be star (Object 2 in the figure)
lies 25 arcseconds from the X-ray position.

Based on the positional coincidence, Object 1 is most probably the
counterpart to the X-ray pulsar, although Object~2 cannot be wholly
dismissed at this stage.

\subsection{AX J0051-722}

This source was first detected as a 91.12 second pulsar in {\it RXTE}
observations (Corbet et al. 1998a) although was initially confused with
the nearby 46 second pulsar 1WGA J0053.8-7226 (Buckley et
al. 1998). Further observations with {\it ASCA} revealed two pulsars
in the field with an approximate 2 to 1 ratio in periods, the 91
second period belonging to the new source, AX J0051-722, whilst
observations with ROSAT reduced the positional uncertainty to 10
arcseconds. We performed spectroscopic observations of the brightest
object in this error circle on 1998 February 3 (see finder chart in
Figure 1(c)). The spectrum (Figure 3(b))
shows the H$\alpha$ line strongly in emission, with an equivalent
width of -22~\AA. The centre of the line corresponds to a velocity of
165$\pm23$~km~s$^{-1}$, consistent with SMC membership.

We have no photometry of objects in this field, but estimate $V \sim
15$ from Digitised Sky Survey images. This, together with the
$H\alpha$ emission and radial velocity indicates an early Be star in
the SMC. With an X-ray positional uncertainty of only 10 arcseconds, we
conclude that this Be star is the optical counterpart to the X-ray pulsar.

\subsection{RX J0051.8-7231}

RX J0051.8-7231 was discovered in {\it Einstein} observations. The
X-ray error circle included the bright SMC star AV~111 Figure 1(d)), which was
suggested as the optical counterpart (Bruhweiler 1987, Wang \& Wu
1992). 

The source displays time variability, as demonstrated by an increase
in luminosity by more than a factor of ten in $\sim$1~year, and by a
factor of $\sim$5 in 5~days (Kahabka \& Pietsch 1996). The
X-ray luminosity, time-variability and hard spectrum led Kahabka \&
Pietsch to suggest a Be/X-ray binary nature for the source. 

Israel et al. (1995) detected pulsations from the X-ray source, with a period
of 8.9 seconds. They subsequently obtained optical
CCD images of the field (Israel et al 1997), and concluded that both
AV111 and a second star (Star 1) within their error circle
displayed H$\alpha$ activity.

In fact, both AV111 and Star 1 do not lie within the error circle
derived by Kahabka \& Pietsch, which contains only one object visible
in the DSS image, star~2. On 1998 February 2 we obtained a medium
resolution H$\alpha$ spectrum of this object, which is shown in Figure
3(c). H$\alpha$ is in emission with EW(H$\alpha$) =
-3.0$\pm$1.0~\AA.

The {\it ROSAT} error circle of Kahabka \& Pietsch seems to
favour the identification of star~2 as the optical counterpart to RX
J0051.8-7231, though in the light of the H$\alpha$ activity detected
by Israel et al. (1997) we cannot rule out star~1, whilst AV111 appears a
less likely candidate.

\subsection{1WGA J0054.9-7226}

%\begin{figure*}
%\begin{center}
%{\epsfxsize 0.9\hsize
%\leavevmode
%\epsffile{0054_bvrh.eps}
%}\end{center}
%\vspace{11cm}
%\caption{{\bf Left:} H$\alpha$ emission index versus $b-v$ colour index for
%objects detected in each of the $BVR$ and $H\alpha$ images of the
%field of the X-ray source 1WGA J0054.9-7226. {\bf Right:} Same diagram
%only for objects within 20 arcseconds (twice the positional
%uncertainty) of the ROSAT X-ray position. Objects that lie within the
%X-ray error circle are plotted as filled circles. The object highest
%up this panel is Star 1.}
%\label{bruh9_bvrh}
%\end{figure*}

\begin{figure}
\begin{center}
{\epsfxsize 0.99\hsize
\leavevmode
\epsffile{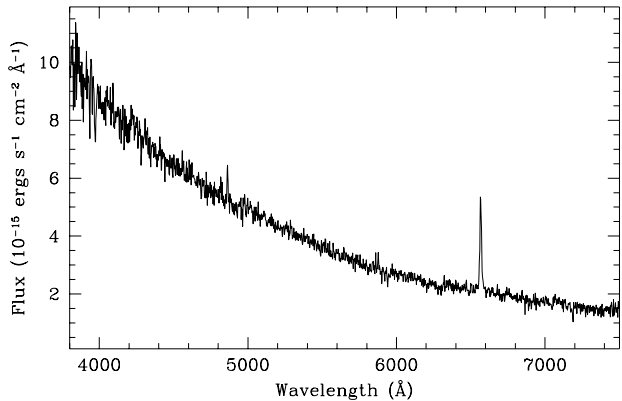}
}\end{center}
%\vspace{11cm}
\caption{Flux calibrated low resolution spectrum of the proposed optical counterpart to the X-ray
source 1WGA J0054.9-7226.}
\label{ma810_low}
\end{figure}

%\begin{table}
%\begin{centering}
%\caption{Photometry of the objects within the X-ray error
%circle of RX J0054.9-7226. Only those objects detected in all of $BVR$
%and H$\alpha$ frames are included.}
%\begin{tabular}{lccccc}
%\hline
%& V & B-V &b-v &R-H$\alpha$\\
%\hline
%1 &14.88$\pm$0.07& -0.13$\pm$0.09& -0.07$\pm$0.09&  0.39$\pm$0.10\\
%2 &15.23$\pm$0.04& -0.10$\pm$0.20&  0.07$\pm$0.19& -0.03$\pm$0.11\\
%3 &16.10$\pm$0.06& -0.19$\pm$0.06&  0.02$\pm$0.06&  0.00$\pm$0.04\\
%4 &16.84$\pm$0.05& -0.08$\pm$0.17& -0.14$\pm$0.13& -0.24$\pm$0.08\\
%\hline
%\end{tabular}
%\label{0049tab}
%\end{centering}
%\end{table}

This source appears in a number of catalogues based on {\em Einstein}
observations of the SMC (Inoue, Koyama \& Tanaka 1983; Bruhweiler et al. 1987; Wang \& Wu 1992),
and was detected by {\it ROSAT} (KP1996). Kahabka and Pietsch (KP1996)
rejected an X-ray binary nature for the source on the grounds of the
lack of time variability (on a timescale of hours).

Observations made with the $RXTE$ satellite on 1998 January 20
detected a 59 second pulsar, with a positional uncertainty of $\pm$10
arcminutes, consistent with the position of RX J0054.9-7226 (Marshall
\& Lochner 1998). Subsequent observations with $SAX$ reduced the
positional uncertainty to a radius of 50 arcseconds, confirming the
identification with RX J0054.9-7226, and refining the pulse period to
58.969 seconds (Santangelo \& Cusumano 1998). The uncertainty in X-ray
position was further reduced to a 10 arcsecond radius from the
analysis of archival ROSAT data by Israel (1998). 

We obtained $BVR$ and H$\alpha$ images of the field on the night of
1996 October 5. An $emission-colour$ diagram for the field revealed 
four objects within the X-ray error circle, all
identified as early type stars by their $B-V$ colours. Of these, only
one shows strong H$\alpha$ emission, with a R-${H\alpha}$ value of
0.49 (the object indicated in Figure 1(e)). 

We obtained spectra of Object~1 on the nights of 1998 February 4 and
5. The medium resolution H$\alpha$ spectrum obtained is shown in
Figure 3(d). The H$\alpha$ line shows strong emission, with
an equivalent width of EW(H$\alpha$) = -25~$\pm$2~\AA, and a radial
velocity of 137~$\pm$28km~s$^{-1}$, consistent with SMC
membership. The low resolution spectrum in Figure~\ref{ma810_low}
shows H$\beta$ also clearly in emission, with EW(H$\beta$) =
-2.0$\pm$1.0~\AA.

\subsection{RX J0101.0-7321}

%\begin{figure}
%\begin{center}
%{\epsfxsize 0.9\hsize
%\leavevmode
%\epsffile{r0101_rh.eps}
%}\end{center}
%\vspace{11cm}
%\caption{H$\alpha$ versus R magnitudes for stars in the field of the 
%X-ray source RX J0101.0-7206.}
%\label{0101hr}
%\end{figure}

This X-ray source was discovered in {\it ROSAT} pointed observations
in 1991 October. Observations approximately 6 months later failed to
detect the source, suggesting a transient nature (Kahabka \& Pietsch
1996). KP1996 claim that the source is most likely associated with a
15-16th magnitude Be star. To the authors' knowledge, no observations
of this star have previously been published. 

We only obtained $R_{C}$ and H$\alpha$ images of the field of RX J0101.0-7321 on
1995 October 1. 
A plot of the measured H$\alpha$ magnitudes 
against the measured $R_{C}$-band magnitudes shows a clear linear relationship
exists between R and H$\alpha$; the
scatter at magnitudes R $>$ 15.5 is mostly due to uncertainties in the
H$\alpha$ magnitudes. Three points show H$\alpha$ excesses which
appear to be much greater than the local scatter of points. Objects~2
and~3 each lie a few arcminutes from the X-ray position, whilst the
error associated with this position is only 11~arcsec. Object~1
is only 10 arcsec from the X-ray position (see the finder chart in
Figure 1(f)).

On 1998 February 3 we obtained a H$\alpha$ spectrum of Object~1. The
spectrum, shown in Figure 3(e), has a strong H$\alpha$
emission line, with EW(H$\alpha$)=-60~\AA. The peak of the H$\alpha$ line is
at a wavelength of 6566.0$\pm$0.5~\AA, corresponding to a
velocity of 148$\pm$23~km~s$^{-1}$. These data confirm that Object~1 is
a Be star in the SMC. As this is the only such object within several
arcminutes of the X-ray position, we identify object~1 as the optical
counterpart, and confirm a Be/X-ray binary nature for this X-ray
source.

\subsection{EXO 0531.1-6609}

%\begin{figure*}
%\begin{center}
%{\epsfxsize 0.9\hsize
%\leavevmode
%\epsffile{0531_bvrh.eps}
%}\end{center}
%\vspace{11cm}
%\caption{{\bf Left:} H$\alpha$ emission index versus $b-v$ colour index for
%objects detected in each of the $BVR$ and $H\alpha$ images of the
%field of the X-ray source EXO 0531.1-6609. {\bf Right:} Same diagram
%only for objects within 27 arcseconds (three times the positional
%uncertainty) of the ROSAT X-ray position. Objects that lie within the
%X-ray error circle are plotted as filled circles.}
%\label{0531_bvrh}
%\end{figure*}

\begin{figure}
\begin{center}
{\epsfxsize 0.9\hsize
\leavevmode
\epsffile{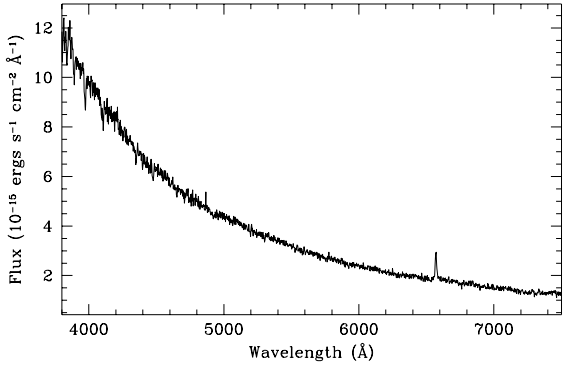}
}\end{center}
%\vspace{11cm}
\caption{Flux calibrated low resolution spectrum of the {\em Northern}
component of the double candidate counterpart to the X-ray
source EXO 0531.1-6609.}
\label{0531_low}
\end{figure}

This source was discovered by EXOSAT during deep observations of the
LMC X-4 region in 1983 (Pakull et al. 1985). It was detected again in
1985 by the SL2 XRT experiment. The lack of detection in EXOSAT
observations made between these dates demonstrates the transient nature
of the source. The object was identified with a Be star by Pakull (private
communication). The counterpart proposed by Pakull is the northern
component of a close double. 

The components of this double are marked 1 and 2 in Figures 1(g). The
positions of the two objects in the {\em emission-colour} diagram show
that both are early type stars. With our chosen criterion of a Be star
having R-H$\alpha \ge 0.2$ then Object~1 is the only Be star within
the X-ray error circle, with R-H$\alpha$ = 0.21, Object~2 has
R-H$\alpha$ = 0.15; within the uncertainties however, our data do not
favour one object over another as emission line objects.

On 1998 February 3 we obtained a spectrum of the Northern component
of the double. The resulting spectrum shown in Figure 3(f)
confirms the presence of H$\alpha$ emission, with EW(H$\alpha$) =
-10.0$\pm$1.0~\AA. The line is centered on a wavelength of 6568.7$\pm$0.5~\AA,
corresponding to a velocity of 272$\pm$23~km~s$^{-1}$. On 1998
February 5 we obtained a low resoluion, flux calibrated spectrum of
this object (shown in Figure~\ref{0531_low}), showing H$\beta$ also
in emission with EW($\beta$) = -0.5$\pm$0.2~\AA.
No spectrum has yet been obtained of the Southern
component of the double.

In order to determine which object is the counterpart to the
X-ray source, it will be necessary to obtain an X-ray position to
sub-arcsecond accuracy, possible with the forthcoming {\it AXAF}
mission, or to find optical/infrared variations in one of the objects
which correlate with X-ray behaviour.

\subsection{H0544-655}

%\begin{figure*}
%\begin{center}
%{\epsfxsize 0.90\hsize
%\leavevmode
%\epsffile{h0544_bvrh.eps}
%}\end{center}
%\vspace{11cm}
%\caption{{\bf Left:} H$\alpha$ emission index versus $b-v$ colour index for
%objects detected in each of the $BVR$ and $H\alpha$ images of the
%field of the X-ray source H0544-665. {\bf Right:} Same diagram
%only for objects within 120 arcseconds (twice the positional
%uncertainty) of the {\it HEAO~1} X-ray position. Objects that lie within the
%X-ray error circle are plotted as filled circles.}
%\label{h0544_bvrh}
%\end{figure*}

\begin{figure}
\begin{center}
{\epsfxsize 0.9\hsize
\leavevmode
\epsffile{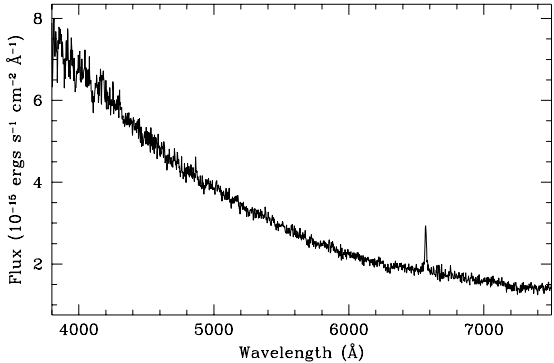}
}\end{center}
%\vspace{11cm}
\caption{Flux calibrated low resolution spectrum of the proposed optical counterpart to the X-ray
source H0544-665.}
\label{h0544_low}
\end{figure}

This source was discovered with the {\it HEAO-1} scanning modulation
collimator by Johnston, Bradt \& Doxsey (1979). The brightest object
within the X-ray error circle (star~1 in Figure~\ref{finders}, and
in Figure~6 of Johnston, Bradt \& Doxsey 1979) was
found to be a variable B0-1 star (van der Klis et al. 1983 and
references therein) but no emission lines have been observed in its
spectrum to identify it as a Be star. van der Klis et al. (1983)
published photometry which showed a negative correlation between
optical magnitudes and colour indices, typical of Be stars whose
variability is due to variations in the circumstellar disc. The
authors expressed concern at the lack of other obvious Be star
spectral characteristics, but suggested that the object may be a Be
star in a low state of activity.

An {\it emission-colour} diagram for objects in the field was created
as previously described. In the 4 arcminute radius area searched, only
one object displays the colours indicative of a Be star. This object
is identified as Object~1 in Figure~\ref{finders}, and corresponds to the
Object~1 of Johnson, Bradt \& Doxey (1979).

We obtained optical spectra of this object in 1998 February, these are
shown in Figures 3(g) and~\ref{h0544_low}. The H$\alpha$ line is
clearly in emission, with an equivalent width measured from the medium
resolution spectrum (Figure 3(g)) of EW(H$\alpha$) =
-8.7$\pm$1.0~\AA. The profile is double peaked with a peak separation
of 181$\pm$30~km~s$^{-1}$, the mean velocity of these peaks is
282$\pm$20~km~s$^{-1}$, consistent with the LMC velocity of
275~km~s$^{-1}$ given by Westerlund (1997), but lower than the
369$\pm$42~km~s$^{-1}$ measured for Balmer absorption lines in this
object by van der Klis et al (1983). The low resolution spectrum in
Figure~\ref{h0544_low} shows also weak H$\beta$ emission with
EW(H$\beta$) = -0.7$\pm$0.5~\AA.

To the authors knowledge, these observations represent the first detection
of emission lines in the spectrum of this star. We confirm a LMC Be
star nature and, due to the lack of other emission line objects in or
near the X-ray error circle, we conclude that this star is the optical
counterpart of the X-ray source.

\section{Discussion}

\begin{figure}
\begin{center}
{\epsfxsize 0.90\hsize
\leavevmode
\epsffile{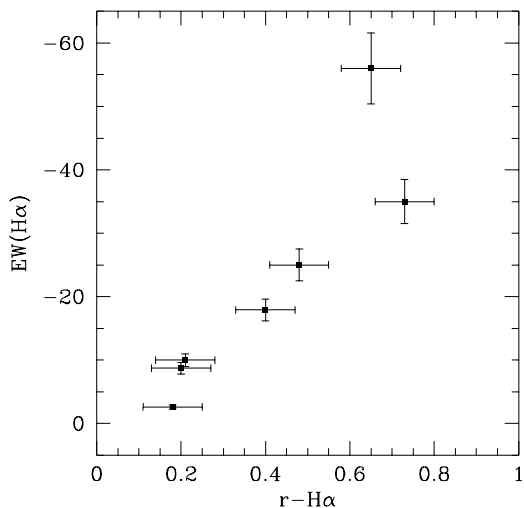}
}\end{center}
%\vspace{11cm}
\caption{Equivalent width of the H$\alpha$ line plotted against the
R-H$\alpha$ emission index. Note that the spectroscopic observations
were made over one year after the photometric observations. The
highest point is that of RX J0101.0-7321.}
\label{ew_rh}
\end{figure}

The number of Be/X-ray binary systems known in the Magellanic Clouds
is now increased to 12 in the SMC and 7 in the LMC. These numbers
compare with 29 known in our own Galaxy. Scaling simply by mass, we
should expect the number of Be/X-ray binaries in the LMC and SMC
to be 0.1 and 0.01 times the number in the Galaxy respectively. In
fact we find, especially in the SMC, what appears at first to be an
abnormally large Be/X-ray binary population.

The discrepancy might be explained through consideration of a number
of issues:

a) Studies of cluster populations in the SMC have shown a
higher proportion of Be stars amongst early type stars than in the
Galaxy; this may be due to the effects of metallicity on a radiatively
driven wind. If a higher proportion of B type stars in the SMC have
circumstellar envelopes, then for a given population of B star/Neutron
star binaries, more SMC systems would be accreting X-ray systems.

b) The star formation history of the SMC may have resulted in a hump
in the stellar age distribution such that a higher proportion of SMC
stars are of the appropriate age to have evolved into Be/X-ray binary
systems. Such a scenario would require an increase in star formation
activity $\sim 10^{7}$ years ago. Supporting evidence for such an
effect comes from the HI work of Staveley-Smith et al (1997) who deduce
from studies of six expanding shells in the SMC a dynamical age of
5.4Myr. They conclude that there must have been an exceptional degree
of coherent star formation throughout the SMC at this time. 

The photometric method used to identify Be star candidates has proved
succesful. In each case where spectroscopic observations have been
made, the Be star nature of the object has been confirmed. The
strength of this approach is illustrated by Figure~\ref{ew_rh} which
shows the relationship between the measured R--H$\alpha$ index from
photometric data and the EW(H$\alpha$) of each object for which both
such data were available. There appears to be a direct correlation,
with the exception of one possible point -- that of RX
J0101.0-7321. This seemingly anomalous point may be explained by
variability, as the photometric and spectroscopic observations were
made over 1 year apart. The indication then is that the Be star
counterpart in this system underwent some change between 1996 October
and 1998 February, which resulted in a large increase in the amount of
circumstellar material. Apart from this kind of inherent difficulty,
this approach is clearly an excellent method for identifying the
counterparts to High Mass X-ray Binaries.

In addition, one must be careful when examining so many fields of the
chance probability of finding a Be star unrelated to the X-ray
source. From the 5 fields presented here, a total of 28 objects were
found lying within twice the error circle radius with an H$\alpha$
index of (R-H$\alpha$)$>$0.2. The total area covered by this sample is
103,000 sq. arcsec. This gives an average Be star rate of almost
exactly 1 per sq. arcmin, so clearly one has to be careful when
working with arc minute size error circles (i.e. RX J0032.9-7348 in
this work). Ultimately, either the error circles have to be reduced by
a significant factor, or simultaneous variability observed between the
X-ray and the optical/IR band, before definite confirmation of the
counterpart is established.

\section{Conclusions}

We have identified several X-ray sources with Be stars in the Small
Magellanic Cloud and demonstrated a reliable technique for doing
so. The exceptionally large number of these systems in the SMC can
probably be attributed to an unusally large amount of star formation 
$\sim 10^{7}$ years ago.

\subsection*{Acknowledgments}

We are grateful to the staff at the SAAO for their assistance, and to
Dr. M. Pakull for communications regarding certain sources. The {\it Shara
Focal Reducer} is on loan from Dr Michael Shara (STScI). JBS
acknowledges the receipt of a Southampton University studentship.

\bsp

\end{document}